\documentclass[conference, letterpaper]{IEEEtran}
\IEEEoverridecommandlockouts

\makeatletter
\def\bstctlcite{\@ifnextchar[{\@bstctlcite}{\@bstctlcite[@auxout]}}
\def\@bstctlcite[#1]#2{\@bsphack
  \@for\@citeb:=#2\do{%
    \edef\@citeb{\expandafter\@firstofone\@citeb}%
    \if@filesw\immediate\write\csname #1\endcsname{\string\citation{\@citeb}}\fi}%
  \@esphack}
\makeatother

\usepackage[utf8]{inputenc}
\usepackage[english]{babel}
\usepackage{cite}
\usepackage{amsmath, amssymb, amsfonts, amsthm}
\usepackage{thmtools, thm-restate}
\usepackage{enumitem}
\usepackage[bb=boondox]{mathalfa}
\usepackage[ruled]{algorithm2e}
\usepackage{graphicx}
\usepackage{subcaption}
\usepackage{textcomp}
\usepackage[dvipsnames]{xcolor}
\usepackage{mathtools}
\usepackage{multirow}
\usepackage{siunitx}
\usepackage{optidef}
\usepackage{nicematrix}
\usepackage{hyperref}

\usepackage[capitalise]{cleveref}

\usepackage{tikz}
\usetikzlibrary{calc}

\usepackage{etoolbox}
\newtoggle{arxiv}
\toggletrue{arxiv}  % arXiv build
\DeclareMathOperator*{\argmin}{argmin}
\DeclareMathOperator*{\argmax}{argmax}

\newcommand{\CVHull}{\mathbf{ConvexHull}}
\newcommand{\W}{\mathbf{W}}
\newcommand{\M}{\mathbf{M}}
\newcommand{\vW}{\vec{W}}
\newcommand{\I}{\mathbf{I}}
\newcommand{\N}{\mathbf{N}}
\newcommand{\x}{\vec{x}}
\newcommand{\xsys}{\vec{\mathbf{x}}}
\newcommand{\xm}{\vec{\mathbf{x}}_{m}}
\newcommand{\R}{\vec{r}}
\newcommand{\Rp}{\vec{\mathbf{r}}_{p}}
\newcommand{\1}{\mathbb{1}}

\declaretheorem{theorem}
\newtheorem{lemma}{Lemma}
\newtheorem{problem}{Problem}

\begin{document}
    \bstctlcite{IEEEexample:BSTcontrol}
    \title{Multiagent Social Influence: \\ \huge{Modeling Persuasion in Contested Social Networks}}

\author{
Renukanandan Tumu\IEEEauthorrefmark{1},
Cristian Ioan Vasile\IEEEauthorrefmark{2},
Victor Preciado\IEEEauthorrefmark{1},
and Rahul Mangharam\IEEEauthorrefmark{1}
\thanks{\IEEEauthorrefmark{1}Department of Electrical and Systems Engineering, University of Pennsylvania, Philadelphia, PA, USA. \{nandant, preciado, rahulm\}@seas.upenn.edu}
\thanks{\IEEEauthorrefmark{2}Department of Mechanical Engineering and Mechanics, Lehigh University, Bethlehem, PA, USA. crv519@lehigh.edu}
\thanks{R. Tumu is supported by the NSF Graduate Research Fellowship Program under Grant No. DGE-2236662.}
}

\maketitle

\begin{abstract}
    We present the Social Influence Game (SIG), a framework for modeling adversarial persuasion in social networks with an arbitrary number of competing players. Our goal is to provide a tractable and interpretable model of contested influence that scales to large systems while capturing the structural leverage points of networks. Each player allocates influence from a fixed budget to steer opinions that evolve under DeGroot dynamics, and we prove that the resulting best-response optimization problem is a difference-of-convex program. To enable scalability, we develop an Iterated Linear (IL) solver that approximates player objectives with linear programs. In experiments on random and archetypical networks, IL achieves solutions within 7\% of nonlinear solvers while being over 10× faster, scaling to large social networks. This paper lays a foundation for asymptotic analysis of contested influence in complex networks.
\end{abstract}
% \IEEEkeywords{}
% \thanks{R. Tumu is supported by the NSF GRFP award DGE-2236662}
    \section{Introduction}
% In democratic societies, opinions matter. The opinions of individuals drives
% their votes, the way they interact with their peers, and the way they interact
% with the world. In recent memory, these opinions have been manipulated by state
% and non-state actors \cite{duncan_gun_2025, kao_infamous_2022}. The interference
% of state actors has caused the annulment of the election results in a NATO member
% state \cite{popescu-zamfir_russian_2025}. In order to better understand these
% influence operations and how they can be countered, we propose the influence game
% setting, based on the extensive study of opinion dynamics.

The contemporary digital landscape is a battleground for influence, where political parties, corporate entities, and state actors compete to shape public opinion. While many study the propagation of misinformation \cite{acemoglu_spread_2010, dave_social_2022}, or the dynamics of political polarization \cite{biondi_dynamics_2023,gaitonde_polarization_2021,proskurnikov_opinion_2016}, they do not fully capture the competitive nature of modern persuasion. Existing work on strategic influence focuses on two-player games \cite{bauso_opinion_2016, de_vos_influencing_2024, lever_strategic_2010, bayiz_effect_2025, bilo_opinion_2016}. However, these frameworks fail to capture the complexity of real-world situations, such as multiparty elections or competing advertising campaigns, where numerous adversarial players operate simultaneously.

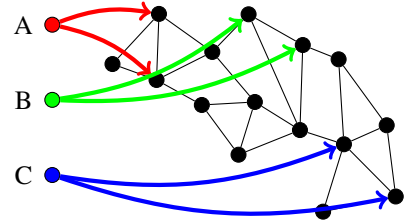
\begin{figure}
    \centering
    \begin{tikzpicture}[%
              scale=2.0,
              every node/.style={circle,draw,fill=black,inner sep=2pt},
              node distance=1cm
            ]

            % Nodes placed at new coordinates
            \node (n0)  at (-0.7071,  0.1323) {};
            \node (n1)  at (-0.6910,  0.5718) {};
            \node (n2)  at (-0.3365,  0.3188) {};
            \node (n3)  at (-0.0952,  0.5692) {};
            \node (n4)  at ( 0.2650,  0.3648) {};
            \node (n5)  at ( 0.5032,  0.2697) {};
            \node (n6)  at ( 0.8828, -0.6428) {};
            \node (n7)  at ( 0.8228, -0.1690) {};
            \node (n8)  at ( 0.5385, -0.2954) {};
            \node (n9)  at ( 0.2462, -0.2009) {};
            \node (n10) at ( 0.4001, -0.7437) {};
            \node (n11) at (-0.0536, -0.0141) {};
            \node (n12) at (-0.1622, -0.3648) {};
            \node (n13) at (-0.4060, -0.0340) {};
            \node (n14) at (-1.0000,  0.2383) {};

            % Example edges (replace with actual graph edges if needed)
            \foreach \i/\j in {0/1,0/2,0/13,0/14,1/2,1/14,2/3,2/11,3/4,3/9,4/5,4/9,5/7,5/8,6/7,6/8,7/8,8/9,9/11,9/12,10/6,10/8,11/12,11/13,12/13} {
                \draw (n\i) -- (n\j);
            }

            % Player nodes on the left side, vertically centered
            \node[fill=red,label=left:A] (R) at (-1.4, 0.5) {};
            \node[fill=green,label=left:B] (G) at (-1.4, 0.0) {};
            \node[fill=blue,label=left:C] (B) at (-1.4,-0.5) {};

            % Influence arrows (thicker)
            \draw[->, red, ultra thick, bend left=15] (R) to (n1);
            \draw[->, red, ultra thick, bend left=15] (R) to (n0);

            \draw[->, green, ultra thick, bend right=15] (G) to (n3);
            \draw[->, green, ultra thick, bend right=15] (G) to (n4);

            \draw[->, blue, ultra thick, bend right=15] (B) to (n6);
            \draw[->, blue, ultra thick, bend right=15] (B) to (n8);

\end{tikzpicture}
    \small{\caption{The Social Influence Game, where players $A$, $B$, and $C$ are competing to influence the opinions held in the social network made up of the black dots, which represent individuals in the social network.
    The colored lines represent allocations of influence from the players to individuals
    in the network.}}
    \label{fig:sig-illustration}
    \vspace{-10pt}
\end{figure}

To address this gap, this paper introduces the Social Influence Game, a novel framework for modeling adversarial persuasion among an arbitrary number of competing players within a social network. Our formulation captures the strategic allocation of influence from fixed budgets, as each player seeks to pull the network's collective opinion toward their own predefined objective (see \cref{fig:sig-illustration}), using DeGroot dynamics \cite{degroot_reaching_1974} for tractability. We formally define this P-player game, prove that synthesizing a best response can be formulated as a Difference of Convex program, and develop an efficient Iterated Linear solution method that scales effectively to larger networks. Numerical studies validate our solver's performance against established benchmarks and provide insights into strategic influence on various archetypal network structures. This work addresses a central question:

\begin{problem}
Given a social network, how can an external agent best allocate a limited influence budget to steer the network's opinion towards a specific goal?
\end{problem}

Our proposed game formulation has two core aspects:
%There are two core aspects of the game formulation we propose; 
(1) The game is adversarial, players compete with each other in order to influence the behavior of the network, (2) The game formulation is multi-player, where the number of players
$P$ can be greater than 2. This $P$-player framework is critical for modeling scenarios like a two-party election where a third-party agent (e.g., a foreign actor or independent PAC) also exerts influence, creating a multi-player ($>2$) contest).

%In order to model even a two-party election with a single external influence agent, our formulation must be able to accommodate more than two players. Our contributions are as follows.

\begin{enumerate}[leftmargin=*]
    \item We formulate the Social Influence Game, a novel framework for analyzing adversarial influence between P players. 
    
    \item We prove that computing a best response can be cast as a Difference of Convex (DC) program, identifying its fundamental mathematical structure.
    
    \item We provide a scalable method to generate influence agent best-responses in the $P$-player social influence game using a Iterated Linear method, which achieves solutions within 7\% of a standard nonlinear solver while being over an order of magnitude faster. 
    \item We analyze the game's outcomes on various network topologies, yielding insights into the strategic value of high-centrality nodes and the opinion-stabilizing effects of large influence budgets.
    
\end{enumerate}

% \subsection{Notation}
\noindent
\textbf{Notation.}
Matrices are denoted by bold uppercase letters ($\mathbf{W}$), and vectors are denoted by lowercase letters with an arrow ($\x_{1}$). 
Tildes ($\mathbf{\tilde{N}_m}$) indicate row-stochastic matrices.
%A matrix will be bolded, like this: . A normalized matrix will have a tilde, like this: . A vector will have an arrow, like this:.
    %\section{Background}
\section{Related Work}
Our work lies at the intersection of opinion dynamics and strategic multi-agent systems. Opinion dynamics models build from the assumption that opinion formation is related to one's relationship with their social connections. Classic opinion models include DeGroot, where one's opinion update as a weighted average of their neighbors \cite{degroot_reaching_1974}, and Friedkin-Johnsen (FJ), which adds individual stubbornness \cite{friedkin_social_1990}. Krause et al. consider bounded confidence, where persuasion only occurs when individuals have similar opinions \cite{ulrich_krause_discrete_2000, hegselmann_opinion_2002}. More contemporary work has sought to capture negative influence \cite{epitropou_opinion_2019}, innovation diffusion \cite{acemoglu_diffusion_2011}, or political polarization \cite{gaitonde_polarization_2021}.

The problem of two-player influence allocation has been studied under FJ \cite{li_robust_2022, shrinate_leveraging_2025}, DeGroot \cite{de_vos_influencing_2024, lever_strategic_2010, grabisch_strategic_2015, liu_two-stage_2025}, and other models \cite{proskurnikov_opinion_2016, bhatt_streisand_2020}. While these papers pose the problem in two-player settings, they focus on different objectives and mechanisms of influence. A common approach is to frame players as attacker and defender, with the attacker seeking to perturb the opinion, and the defender seeking to prevent this \cite{li_robust_2022, bayiz_effect_2025, grabisch_strategic_2015}. The available mechanisms vary, where some consider that a strategic agent is only able to make a single connection \cite{grabisch_strategic_2015}, others allow the addition of links and the modification of influence weights \cite{li_robust_2022}, and still others allow symmetric message passing between attacker and defender \cite{bayiz_effect_2025}. Contagion games have been used to model counterinsurgency efforts, and the allocation of limited budget to select individuals \cite{tsai_security_2012}. A similar influence maximization game occurs in marketing \cite{goyal_competitive_2012, bharathi_competitive_2007, fazeli_game_2012}. 

The formulation of games on social networks has also been used to capture opinion formation. Bauso et al. model consensus and dissensus through a Mean Field Game \cite{bauso_opinion_2016}, while Bhatt et al. use a contagion process to study the Streisand effect of censorship \cite{bhatt_streisand_2020}. Bindel et al. examine diversity of opinions from the tradeoff between individual conviction and conformity \cite{bindel_how_2015}, and Bilò et al. extend this to dynamic network topologies \cite{bilo_opinion_2016}. In these papers, players are internal to the network and do not aim to shift the overall collective opinion. By contrast, we study external players with adversarial objectives. Christia et al. pose a related multi-player influence problem where agents allocate budgets to adjust initial opinions before DeGroot-style updates evolve the network \cite{christia_scalable_2021}. Their convex setting yields polynomial-time Nash equilibria. Our formulation differs in that external influence is applied at every timestep alongside neighbor averaging, which leads to a non-convex optimization problem. Put differently, Christia et al. study seeding of initial opinions, whereas we study persistent influence.

A key challenge in modeling modern information ecosystems is the simultaneous presence of internal and external actors impacting opinion formation. The primary contribution of our work is to address this gap by proposing a formal P-player game that moves beyond the traditional dyadic conflict model. Our formulation allows for an arbitrary number of competitors, providing a more realistic and generalizable framework for analyzing adversarial persuasion in complex social networks.

% The second focuses on strategic and adversarial interaction, and explores how to formulate the interaction between actors, and modeling or investigating the impact of specific factors. For example, Bhatt et. al. \cite{bhatt_streisand_2020} use the FJ model to capture the Streisand effect in social networks with censor. Others study polarization \cite{biondi_dynamics_2023} and misinformation \cite{acemoglu_spread_2010}, seeking to identify the conditions under which these phenomena develop and the impact of stubborn agents on misinformation, respectively. Other work examines the development of opinions as a game or control problem, where an external agent seeks to manipulate the opinions in the network. Some pose this problem as a two-player game \cite{epitropou_opinion_2019, de_vos_influencing_2024}, where external agents seek to drive opinion toward their preference. Bauso et al. describe a model based on a Mean Field Game in order to capture consensus and dissensus behavior in the network, however, the network level game is still a contest between two players \cite{bauso_opinion_2016}. Bayiz et. al. similarly develop a model based on a bi-level Stackelberg game played between an attacker and a defender, and investigate the impact of the network topology on the resulting equilibrium \cite{bayiz_effect_2025}.

% Our work builds upon the rich literature of opinion dynamics models \cite{degroot_reaching_1974,friedkin_social_1990,ulrich_krause_discrete_2000} and opinion
% influence games \cite{}. 
\textbf{The DeGroot model} is one of the most widely used opinion dynamics
models, and is the basis for our work \cite{degroot_reaching_1974} due to its mathematical tractability. The DeGroot model describes the change in opinion as one of weighted averaging, where the opinion of an individual in the next timestep is given by the trust-weighted
average of their neighbors. The opinions of all individuals are collected in the
opinion vector $\x$. Each opinion is in $\mathbb{R}$. The trust
is aggregated in the matrix $\tilde{\W}$, where $\tilde{\W}_{ij}$ represents the trust placed by
individual $i$ in individual $j$. The trust matrix $\tilde{\W}$ must be row-stochastic,
that is, that the sum of each row of the matrix must be equal to $1$, and all
elements must be positive. 
% Assembled together, the DeGroot dynamics are
% presented in~\cref{eq:degroot}.

% \begin{equation}
% \label{eq:degroot}
%     \x(t+1) = \tilde{\W} \x(t)
% \end{equation}

In this paper, we use DeGroot dynamics for modeling of multidimensional opinions.
For a $D$-dimensional opinion, in a system with $M$ individuals, we get the system state $\xsys(t) = [\x
_{1}^{\top}, \ldots, \x_{M}^{\top}]^{\top}$, and the system dynamics
%
% in \cref{eq:degroot-ext}.
%
% In this paper, we use DeGroot dynamics to model the social network's opinions.
% We extend the DeGroot model to handle multi-dimensional opinions
% by using the Kronecker product.
% For a $k$-dimensional opinion, we get the system state $\vec{x}(t) = [\x
% _{1}^{\top}, \ldots, \x_{M}^{\top}]^{\top}$, and the system dynamics in
% \cref{eq:degroot-ext}.
%
\begin{equation}
    \xsys(t+1) = \left(\tilde{\W}\otimes\I_{D}\right) \xsys(t), \label{eq:degroot-ext}
\end{equation}
where $\otimes$ is the Kronecker product.

It is important to note here that the trust matrix $\tilde{\W}$ here is identical
to the original DeGroot dynamics, simply duplicated across each dimension of
the opinion.

%The primary insight is that the presence of many actors in the modern information ecosystem, who all seek to influence our opinion. At any given time, economic, social, and political actors may all seek to influence the opinions of a social network.

    \section{The $P$-player Influence Setting}

We describe the dynamics of the $P$-player social network with influence here.
We use the extended DeGroot model in \cref{eq:degroot-ext} in order to
accommodate many players while maintaining parity in the resulting game. The social network has two types of participants, \emph{individuals} who comprise the social network and \emph{players} who seek to influence the opinions of the individuals.

\paragraph{Opinion Vectors and the Social Network State}
Each individual in the social network has an opinion vector $\x_{i}\in \mathbb{R}^{D}$.
We consider the aggregated opinions of the individuals in the network to be $\xm$, which is the concatenation of the opinions of each individual $i \in \{1, \ldots, M\}$.
This state is time varying, and $\xm(t)$, and
$\x_{i}(t)$ represent the complete state of the social network and the state of
individual $i$ at time $t$ respectively. Each individual opinion is a $D$ dimensional
vector, and $\xm$ is in $M\cdot D$ dimensions.

Each player $p$ has a reference opinion $\R_{p} \in \mathbb{R}^D$, which is specified before the game
begins. We denote the aggregation of the reference opinions $\Rp$. The player is modeled
as having the opinion $\R_{p}$, with all trust placed on themselves. This ensures
that the player has a static opinion across timesteps. Generating competitive starting
reference opinions is discussed in more detail in \cref{sec:ref-obj}.

The complete system state is the concatenation of the states of all of the
players, followed by the state of all of the individuals in the social network.
% The complete state is presented below in \cref{eq:state-def}.
\begin{equation*}
    % \label{eq:state-def}
    \xsys(t) = \left[\R_{1}^{\top}, \ldots, \R_{P}^{\top}, \x_{1}(t)^{\top},\ldots,\x
    _{M}(t)^{\top} \right]^{\top} = [\Rp^{\top}, \xm^{\top}(t)]^{\top}
\end{equation*}

\paragraph{Trust Matrix Decomposition}
The state changes over time with the trust matrix $\W$. This matrix contains
both the trust matrix of the social network of individuals $\tilde{\W}_{m}$, as well as the
influences allocated by the players. For notational convenience, we call the expanded matrix
${\W}_{m}= \tilde{\W}_{m}\otimes \I_{D}$. Each player chooses their influence
vector $\vW_{p} \in \mathbb{R}^M$ such that each entry is non-negative, and all entries sum to less
than $\lambda$, the influence budget of the agents. These influence vectors are aggregated
into the matrix $\mathbf{W}^{*}=\begin{bNiceArray}{c|c|c}\vW_1 & \hdots & \vW_P \end{bNiceArray} \in \mathbb{R}^{M \times P}$. The assembled trust matrix is 

%$\W^{*}$ as shown in \cref{eq:def-w-star}.
%
% \begin{equation*}
    % \mathbf{W}^{*}=
    % \begin{bNiceArray}{c|c|c}
    %     \vW_1 & \hdots & \vW_P
    % \end{bNiceArray} \in \mathbb{R}^{M \times P}
    % \label{eq:def-w-star}
% \end{equation*}

% The assembled trust matrix $\W$ also contains the identity matrix tiled in the
% upper left, which preserves the reference opinions across time. The trust matrix
% is presented in \cref{eq:def-w}

\begin{equation*}
    \W =
    \begin{bNiceArray}{c|c}
        \I_P & \Block{1-1}{\mathbf{0}} \\
        \hline
        \W^* & \tilde{\W}_m
    \end{bNiceArray}
    % \in \mathbb{R}^{(M+P)D \times (M+P)D}
    % \label{eq:def-w}
\end{equation*}
and contains the identity matrix tiled in the
upper left, which preserves the reference opinions across time.

%The assembled trust matrix in Equation (3) is not row-stochastic because the influence allocations in $\tilde{\W}$ add weight to rows that already sum to one. To remedy this, we introduce a normalization matrix N that restores the row-stochastic property essential for DeGroot dynamics.

While $\tilde{\W}_m$ is row-stochastic, the additional influence from the players in $\W^*$ increase the row sum. Thus, we use the normalization matrix $\mathbf{N}$ in \cref{eq:def-n-mat} to ensure that our assembled $\W$ is row-stochastic. The upper left block is the identity and does not require normalization. The bottom right block is shown in \cref{eq:def-n-middle}.
% which when
% multiplied with $\W$, will result in a row-stochastic matrix.
% The block representation of the $\W$ matrix helps us define the normalization matrix in \cref{eq:def-n-mat}.
% Because the upper left block is the identity, it does not require normalization.
% Only the bottom right block requires normalization, and the size of that normalization
% is shown in \cref{eq:def-n-middle}.
Since $\tilde{\W}_{m}$ is row-stochastic, $\tilde{\W}_{m} \mathbb{1}_{M} = \mathbb{1}_{M}$.

% Because $\tilde{\W}_{m}$ is already row-stochastic,
% $\mathbb{1}_{M}$ is a left-eigenvector, i.e., $\tilde{\W}_{m} \mathbb{1}_{M} = \mathbb{1}_{M}$.
%we add $1$ to the row sum of the influence matrix.
%
\begin{gather}
    \mathbf{N}=
    \begin{bNiceArray}{c|c}
        \I_{P}                  & \Block{1-1}{\mathbf{0}} \\
        \hline
        \Block{1-1}{\mathbf{0}} & \mathbf{N}_m
    \end{bNiceArray}\label{eq:def-n-mat}\\
    \mathbf{N}_{m}= \textrm{diag}\left(\mathbb{1}_{M}+ \mathbf{W}^{*}\mathbb{1}
    _{M}\right)^{-1}\label{eq:def-n-middle}
\end{gather}

The overall system dynamics are $\xsys(t+1) = (\mathbf{N}\W\otimes \I_{D}) \xsys(t)$. Multiplication results in the update equation
% in \cref{eq:combined-update}.
\begin{equation*}
    \begin{bNiceArray}{c}
        \Rp        \\
        \hline
        \xm(t+1)
    \end{bNiceArray}
    = \left(
    \begin{bNiceArray}{c|c}
        \I_P                     & \Block{1-1}{\mathbf{0}}           \\
        \hline
        \mathbf{N}_m\mathbf{W}^* & \mathbf{N}_m \tilde{\mathbf{W}}_m
    \end{bNiceArray}\otimes \I_{D}\right)
    \begin{bNiceArray}{c}
        \Rp      \\
        \hline
        \xm(t)
    \end{bNiceArray}
    \label{eq:combined-update}
\end{equation*}

\paragraph{Asymptotic State Calculation}
While the equation provides the update for the entire
system including the players, we only consider the state of the social network $\xm$
to specify the game objective.

{\small
\begin{align}
    \xm(t+1) & = \left(\mathbf{N}_{m}\mathbf{W}^{*}\otimes \I_{D}\right) \Rp + \left(\mathbf{N}_{m}\tilde{\W}_{m}\otimes \I_{D}\right)\xm(t) 
    \label{eq:middle-update}
\end{align}}%
We consider the asymptotic opinion of the network
% , which is obtained by assuming
% that
when the system reaches a steady state as $t \to \infty$.
% The asymptotic state is 
%presented in \cref{eq:asymp-update}
\begin{equation*}
    \xm(\infty) = \left((\I_M- \mathbf{N}_{m}\tilde{\W}_{m})^{-1}\mathbf{N}_{m}\mathbf{W}^{*}\otimes \I_{D}\right)\Rp %\label{eq:asymp-update}
\end{equation*}
% \begin{align}
%     % \xm(\infty)   & = \mathbf{N}_{m}\mathbf{W}^{*}\R+ \mathbf{N}_{m}\tilde{\mathbf{W}}\xm(\infty) \notag\\
%     % (\I- \mathbf{N}_{m}\tilde{\mathbf{W}})\xm(\infty)     & = \mathbf{N}_{m}\mathbf{W}^{*}\R \notag\\
%     \xm(\infty) & = (\I- \mathbf{N}_{m}\tilde{\W}_{m})^{-1}\mathbf{N}_{m}\mathbf{W}^{*}\otimes \I_{D}\Rp \label{eq:asymp-update}
% \end{align}
The stubborn nature of the player opinions eliminates the influence of the initial individual opinions in the asymptote. To isolate the contribution of individual players, we expand $\W^{*}$ and use the mixed product property of the Kronecker product
% $(\vW_j \otimes I) \cdot (1 \otimes \R_j) = (\vW_j \otimes \R_j)$
%obtain \cref{eq:penultimate-asymp-update}.
\begin{equation*}
    \xm(\infty) = \left(\left((\I- \mathbf{N}_{m}\tilde{\mathbf{W}}_{m})^{-1}\mathbf{N}
    _{m}\right)\otimes \I_{D} \right) \Bigg( \sum_{j=1}^{P}\vW_{j}\otimes\R_{j}\Bigg) \label{eq:penultimate-asymp-update}
\end{equation*}

To simplify this further, and avoid decision variables with division, we define $\N
_{i}$, the inverse of $\N_{m}$, shown in \cref{eq:def-ni}, to obtain the
expression for the asymptotic state \cref{eq:final-asymp-update}.
\begin{gather}
    \N_i= \textrm{diag}\left(\mathbb{1}_{M}+ \sum^{P}_{p=1}\vW_{p}\right)
    = \N_m^{-1}\label{eq:def-ni}\\
    \xm(\infty) = \left((\N_{i}- \tilde{\W}_{m})^{-1}\otimes \I_{D} \right)\left( \sum
    _{j=1}^{P}\vW_{j}\otimes\R_{j}\right)\label{eq:final-asymp-update}
\end{gather}

Moreover, the $\N_{i}- \tilde{\W}_{m}= (\mathrm{diag}(\sum^{P}_{p=1}\vW_{p}+ \mathbb{1}
_{M})) - \tilde{\W}_{m}= (\mathrm{diag}(\sum^{P}_{p=1}\vW_{p}) + \mathbf{L}_{m})$,
where $\mathbf{L}_{m}$ is the graph Laplacian of the social network.
    \section{Social Influence Game}
The players in the game must choose an influence vector that minimizes the distance
between their reference opinion $\R_{p}$ and the mean opinion of the social
network. They are subject to constraints, that the sum of their influence vector
be lower than their influence budget $\lambda$, and that each element of their influence
vector be non-negative. In the following subsections, we present the game objective,
the optimization problem we solve in order to obtain the influence agent
policies, and case studies to demonstrate that the problem is challenging.

\subsection{Objective}
In a network of individuals, we frame the problem of driving the average opinion to the reference opinion, and we seek to minimize the norm of the two quantities, as in \cref{eq:norm-prob-formulation}.

\begin{subequations}
    \label{eq:norm-prob-formulation}
    \begin{align}
        \argmin_{\vW_p}\; & \left\|\R_{p} - \frac{1}{M} \left(\mathbb{1}_{M}^{\top}\otimes \I_{D}\right) \xm(\infty)\right\|_2 \label{eq:orig-objective} \\
        \textrm{s.t.}\;   & \; \vW_{p}\geq 0                                                                                            \\
                          & \; \mathbb{1}_{M}^{T}\vW_{p}\leq \lambda
    \end{align}
\end{subequations}
Although this problem may appear to be a quadratic program, it is not, due to the non-linear structure of \cref{eq:middle-update}. We can simplify this objective by making the assumption that the reference opinions are all unique and of unit norm, and that all initial opinions are within this convex hull. Under these conditions, we show that this minimizer of this norm is equivalent to the maximizer of a dot product, presented formally in \cref{theorem:objective-equality}. In a two-player setting, individual opinions are on the line segment between the two players.

\begin{restatable}[Objective Equivalence]{theorem}{objequal}
    \label{theorem:objective-equality} 
    Given a set of reference opinions $\mathcal{R}= \{\R_{1}, \ldots, \R_{P}\}$ which have the properties that the reference opinions all have the unit norm $\|\R_p\|_2 = 1$, and that reference opinions are distinct $\R_i \neq \R_j$ for all $i\neq j$, and a DeGroot social network of $M$ individuals with starting opinions $\x(0) = \{\x_{k}(0) \in \CVHull(\mathcal{R}), \; \forall \; i\in M\}$, the minimizer of the objective function $J_1(\R_p, \x(t)) = \|\R_p - (\sum_{k=1}^{M} \x_k(t) )/{M}\|_2^2$ maximizes $J_2(\R_p, \x(t)) = \R_p^\top\left(\mathbb{1}_{M}^{\top}\otimes \I_{D}\right) \xm(t)$.
\end{restatable}

This equivalence stems from the stability property of DeGroot dynamics (\cref{lemma:opinion-bounds}). Because opinions are confined to the convex hull of the reference opinions, minimizing the Euclidean distance to a reference opinion $\R_p$ is equivalent to maximizing the projection of the network's opinion vector onto $\R_p$.

%This property can be thought of as a consequence of the stability properties of the DeGroot update models (stated in \cref{lemma:opinion-bounds}. Because the opinions will never leave the convex hull of the reference opinions, the dot products will never exceed a fixed sum and thus will be maximized by the same opinions that minimize the norm. This results in the formulation of the problem in \cref{prob:prob-formulation}.

\begin{problem}
    [Player $p$'s Objective in the Social Influence Game]\label{prob:prob-formulation}
    The objective of the player $p$ is to maximize the dot product of the
    asymptotic opinion of the social network $\xm(\infty)$ and the reference
    opinion $\R_{p}$.
    \begin{subequations}
        \label{eq:prob-formulation}
        \begin{align}
            \argmax_{\vW_p}\; & \R_{p}^{\top}\left(\mathbb{1}_{M}^{\top}\otimes \I_{D}\right) \xm(\infty) \label{eq:orig-objective} \\
            \textrm{s.t.}\;   & \; \vW_{p}\geq 0                                                                                            \\
                              & \; \mathbb{1}_{M}^{T}\vW_{p}\leq \lambda
        \end{align}
    \end{subequations}
\end{problem}

\subsection{Difference-of-Convex Programming}
\label{sec:dc-programming} The problem in \cref{prob:prob-formulation} is a non-convex
optimization problem, as the objective is a non-linear function of the influence
vector $\vW_{p}$. Reformulation of this problem will show that the problem can
be posed as a Difference-of-Convex (DC) programming problem, which is a class of
non-convex optimization problems \cite{horst1999dc} that can be solved efficiently using specialized
algorithms \cite{shen2016disciplinedconvexconcaveprogramming}.

\begin{theorem}
    \label{thm:dc-programming} \cref{prob:prob-formulation} is a Difference-of-Convex
    (DC) program.
\end{theorem}
\begin{IEEEproof}
    The objective in \cref{prob:prob-formulation} is a non-linear function of the
    influence vector $\vW_{p}$. We expand the objective in
    \cref{eq:prob-formulation} using the shorthand
    $\hat{\N}_{i}= \N_{i}\otimes \I_{D}$ and
    $\hat{\W}_{m}= \tilde{\W}_{m}\otimes \I_{D}$ to get:
    \begin{subequations}\small
        \label{eq:dc-initial-expansion}
        \begin{align}
            \argmax_{\vW_p}\; & \R_{p}^{\top}\left(\mathbb{1}_{M}^{\top}\otimes \I_{D}\right)\left(\hat{\M}^{-1}\right) \delta \\
            \textrm{s.t.}\;   & \; \hat{\mathbf{M}}= (\hat{\N}_{i}- \hat{\W}_{m})                                              \\
                              & \; \delta = \sum_{j=1}^{P}\vW_{j}\otimes\R_{j}\label{eq:delta-def}                             \\
                              & \; s = \sum_{j=1}^{P}\vW_{j}                                                                   \\
                              & \; \vW_{p}\geq 0                                                                               \\
                              & \; \mathbb{1}_{M}^{T}\vW_{p}\leq \lambda
        \end{align}
    \end{subequations}
    We define $z^{\top}= \R_{p}^{\top}(\1_{M}^{\top}\otimes \I_{D})\hat{\M}^{-1}$.
    Rearranging terms results in the equation
    $\hat{\M}^{\top}z = \1_{M}\otimes \R_{p}$. Further expansion of these terms results
    in $(\hat{\N}_{i}- \hat{\W}_{m})^{\top}z -\1_{M}\otimes \R_{p}= 0$. The $\%$ operator is defined to be the $\mod$ operator.
    Expanding the $k$-th row, we obtain $(1+s_{k\%D})z_{k}- (\hat{\W}_{m}^{\top})_{k}z - (\R_{p})_{k\%D} = 0$.
    Using the binomial expansion $2ab = (a+b)^{2}- a^{2}-b^{2}$:
    \begingroup
    \begin{equation}\label{eq:dc-constraint}\small
        \begin{split}
            g_{k}(\vW_{p},z) =&\bigl((2+s_{k\%D}+ z_{k})^{2}\\[1pt]
            &\qquad+ \Bigl(2(\hat{\W}_{m}^{\top})_{k}z + 2(\R_{p})_{k\%D}\Bigr)^{2}
            \Bigr) \\
            &\qquad - \bigl( (2 + 2s_{k\%D})^{2}+ z_{k}^{2}+ \\
            &\qquad\qquad 4((\hat{\W}_{m}^{\top})_{k}z)^{2}+ 4((\R_{p})_{k\%D})^{2}
            \bigr)
        \end{split}
    \end{equation}
    \endgroup
    The constraints $g_{k}(\vW_{p},z)$ are DC, as they are a difference of convex
    functions. In order to enforce equality, we set $g_{k}(\vW_{p},z)\leq0$ and $-
    g_{k}(\vW_{p},z)\leq 0$. We rewrite the objective in \cref{eq:dc-initial-expansion} as $z^{\top}\delta$, which
    leads to
    % can be rewritten using the binomial expansion as:
    \begin{equation}
        \begin{split}
            f(\vW_{p},z) =&\frac{1}{2}\left( \|z+ \delta\|^{2}- \left(\|z\|^{2}+\|
            \delta\|^{2}\right) \right)
        \end{split}
    \end{equation}
    This is a DC objective. Therefore, the problem in
    \cref{eq:dc-initial-expansion} is a DC programming problem, and can be written:
    
    \begin{subequations}\small
        \label{eq:dc-initial-expansion}
        \begin{align}
            \argmax_{\vW_p, z}\; & f(\vW_{p}, z)                                  &                    \\
            \textrm{s.t.}\;      & \; g_{k}(\vW_{p}, z) \leq 0                    & \; (k=1,\ldots,Md) \\
                                 & \; -g_{k}(\vW_{p}, z) \leq 0                   & \; (k=1,\ldots,Md) \\
                                 & \; \delta = \sum_{j=1}^{P}\vW_{j}\otimes\R_{j} &                    \\
                                 & \; s = \sum_{j=1}^{P}\vW_{j}                   &                    \\
                                 & \; \vW_{p}\geq 0                               &                    \\
                                 & \; \mathbb{1}_{M}^{T}\vW_{p}\leq \lambda       &
        \end{align}
    \end{subequations}
\end{IEEEproof}

\iftoggle{arxiv}{
\begin{figure*}[bt]
    \centering
    \includegraphics[width=0.8\linewidth]{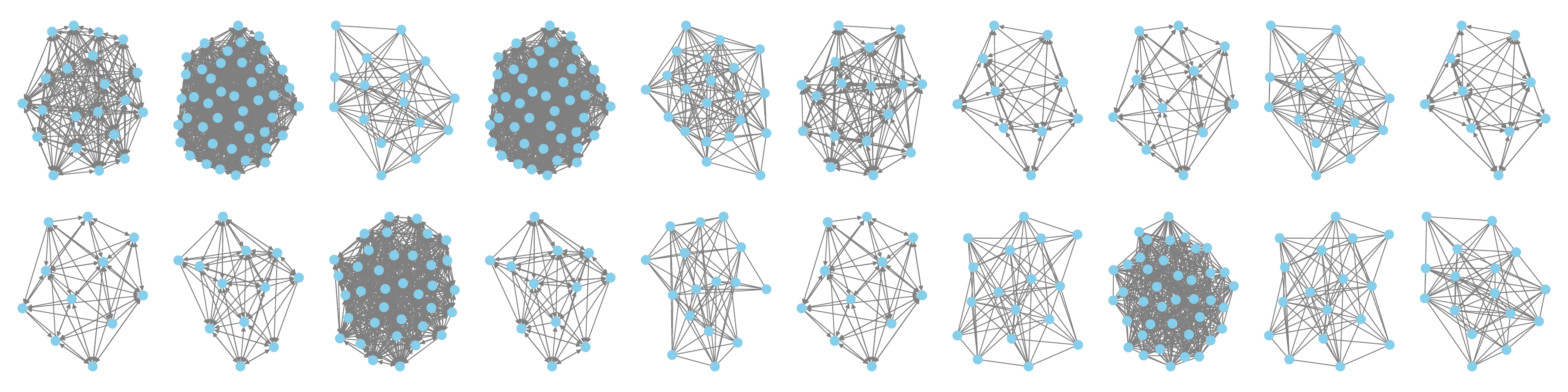}
     \vspace{-10pt}
    \small{\caption{The figure shows network topologies generated to evaluate the performance of the solver. The existence of an edge is modeled as a Bernoulli random variable, and the edge weights are randomized in order to obtain a new network.}}
    \label{fig:example-topologies}
     \vspace{-10pt}
\end{figure*}
}{}

\subsection{Iterated Linear Solution}\label{sec:iterated-linear}
To develop a more scalable solver, we propose an iterative linear approximation. In each iteration, we assume the influence allocations of other players are fixed. This assumption allows us to treat the complex normalization term $\N_{i}$ as a constant matrix, simplifying the objective to a linear function of player p's influence vector, $\vW_{p}$. We can write the objective from \cref{eq:orig-objective} using our definition of
$\delta$ from \cref{eq:delta-def} as:
\begin{equation}
    \underbrace{\R_{p}^{\top}\left(\mathbb{1}_{M}^{\top}\otimes \I_{D}\right)\left(\N_{i}-
    \tilde{\W}_{m}\right)^{-1}\otimes \I_{D}}_{A}\delta
\end{equation}
%The key assumption we make is that the change in the influence vector $\vW_{p}$ is small enough so that the change in $\N_{i}$ is negligible. 
\begin{figure}
    \centering
    \includegraphics[width=0.90\linewidth]{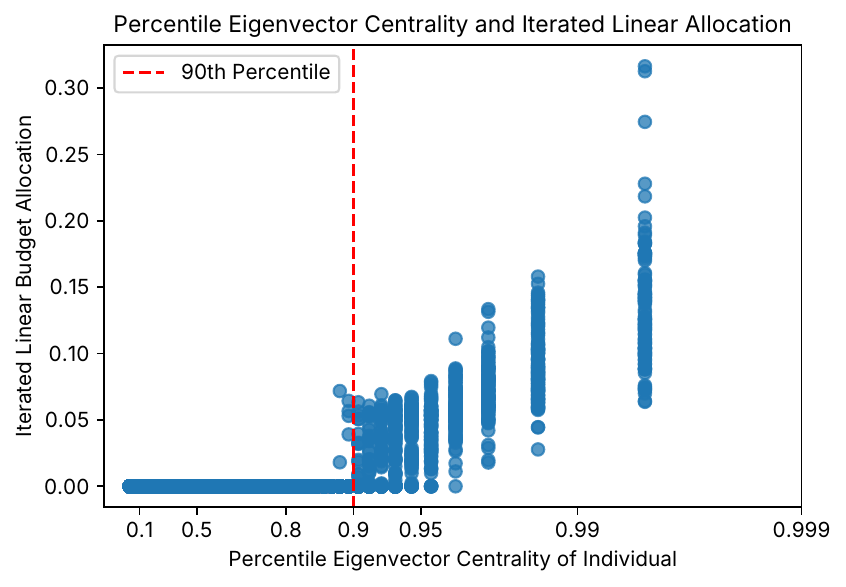}
    \small{\caption{The relationship between eigenvector centrality (percentile) and the budget allocations produced by the iterated linear approach follow a hinged pattern, with allocations remaining flat until the eigenvector centrality is in the top decile.}\label{fig:sol-vs-centrality}}
    \vspace{-20pt}
\end{figure}

This allows us to treat
$A$ as a constant matrix and write the objective as a linear function. With further
manipulation, we can write the objective as:
\begin{equation}
    A\Bigg(\sum_{\substack{j=1 \\ j\neq p}}^{P}\vW_{j}\otimes \R_{j}\Bigg)+ A\left
    (\vW_{p}\otimes \R_{p}\right)
\end{equation}
The left-hand side is constant with respect to $\vW_{p}$, and so we can drop it
from the objective. Applying the mixed-product property of the Kronecker product,
we can rewrite the objective as:
\begin{equation}
    A\left(\I_{M}\otimes \R_{p}\right)\vW_{p}
\end{equation}
This is a linear function of $\vW_{p}$, so we can write the problem as a
linear program:
\begin{subequations}
    \label{eq:iterated-lp}
    \begin{align}
        \argmax_{\vW_p}\; & A\left(\I_{M}\otimes \R_{p}\right)\vW_{p} \\
        \textrm{s.t.}\;   & \; \vW_{p}\geq 0                          \\
                          & \; \mathbb{1}_{M}^{T}\vW_{p}\leq \lambda
    \end{align}
\end{subequations}
We can solve the problem in \cref{eq:iterated-lp} iteratively, updating the
influence vector $\vW_{p}$ at each iteration. Paired with a step size to
ensure that the change in $\vW_{p}$ is small, this method can be used to find a
local optimum of the original problem in \cref{prob:prob-formulation}. We used Nesterov's accelerated gradient descent \cite{nesterov_method_1983} to update our guess of $\vW_{p}$.

We sampled $100$ random networks with $M=100$ individuals each according to the Stochastic Block model \cite{holland_stochastic_1983}, and solved them according to the procedure described in this section. The influence budget $\lambda=0.5$. We compared the influence budget allocations with the eigenvector centrality of each individual. In order to normalize for different network topologies, we plot the relationship between the percentile of the eigenvector centrality of an individual in their social network with the budget allocation generated by the Iterated Linear solver in \cref{fig:sol-vs-centrality}. This reveals a phenomenon where the top percentiles of individuals in centrality receive most of the influence budget. When allocations are made directly according to the eigenvector centrality, the objective value achieved ($0.423 \pm 0.015$) is lower than that achieved by the Iterated Linear approach ($0.500 \pm 0.031$), indicating that this ``spiky" allocation pattern performs better.

% This method is shown to converge in practice, but we do not have a formal proof of convergence.
    \section{Experimental Evaluation}
\label{sec:experiments} 
Our experiments show the performance of our proposed solution compared to several baselines, and the relationship of the solutions as the number of individuals increases. We will present case studies showing the optimal actions for specific types of networks, and we will finally show the relationship between influence budget and average opinion change.

In this section, we present a numerical evaluation of our proposed approach to the Social Influence Game. In our evaluations, we use our Iterated Linear (IL) solver, a Non-Linear Program solver (NLP) \cite{andersson_casadi_2019,wachter_implementation_2006}, a genetic solver (CMA-ES) \cite{hansen_cma-espycma_2023}, and a DCCP solver \cite{shen2016disciplinedconvexconcaveprogramming}. We evaluate the performance of these solvers when the scale of the network changes, show some qualitative results on archetypical networks, and characterize the impact of the influence budget $\lambda$ on the change in average opinion.

\subsection{Reference Objectives}
\label{sec:ref-obj} In our model, each player is assigned a reference opinion, which serves as the target toward which the corresponding influence agent seeks to steer the aggregate opinion of the social network. We aim to generate a set of reference opinions that are unbiased, equidistant, and are in a Euclidean space. By unbiased, we mean that progress towards one player's reference opinion does not produce an advantage to others. A natural choice is to select the vertices of a regular $n$-simplex to represent the reference opinions. This generates a line for two players, an equilateral triangle for three players, and a regular tetrahedron for four players. All reference opinions are unit norm.

\subsection{Solver Performance}
\label{sec:solver-perf} 
\begin{table}[t]
\centering
\begin{tabular}{l
                S[table-format=1.3]
                S[table-format=1.3]
                S[table-format=1.3]
                S[table-format=2.3]}

 & {\shortstack{\textbf{Iterated Linear} \\ \textbf{(Ours)}}} & {NLP} & {DCCP} & {Genetic} \\
\hline
Average Improvement & 0.425 & 0.454 & 0.369 & -2.958 \\
Standard Deviation  & 0.014 & 0.010 & 0.208 &  8.973 \\
\hline

\end{tabular}
\caption{Solver Performance over sampled scenarios.\label{tab:solver-perf}}
\end{table}
\begin{figure}[]
    \centering
    \includegraphics[width=0.9\linewidth]{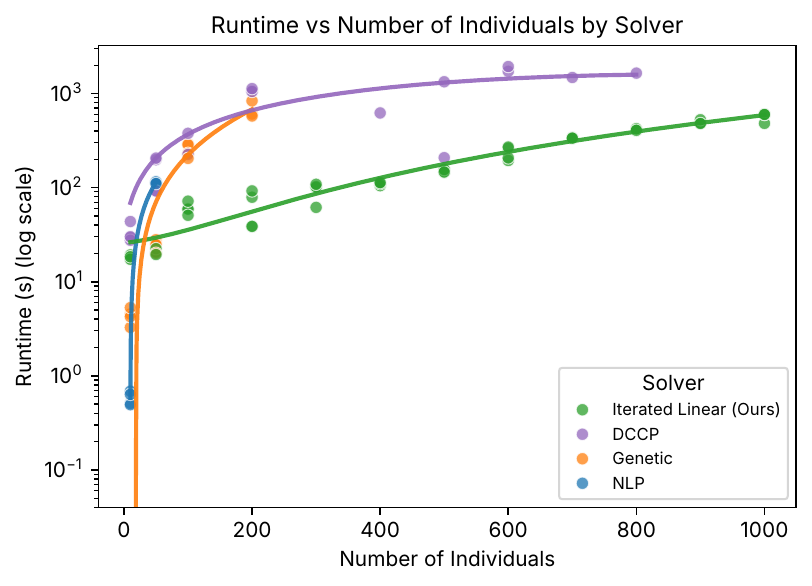}
     \vspace{-10pt}
     \small{\caption{This plot shows the time taken to solve our optimization problem as the number of individuals in the social networks increases. Only the Iterated Linear solver was able to produce a solution when $M\geq900$ within the computation budget of $2000$ seconds.}\label{fig:solver-runtime}}
%    \caption{Solver performance comparison. The top plot shows that the improvement provided by our method, the Iterated Linear solver, is better than the genetic solver. The bottom plot shows the scaling of runtime as the number of individuals increases. The runtime offered by our approach is lower than all baselines when network size exceeds $40$.}
     \vspace{-10pt}
\end{figure}

\iftoggle{arxiv}{
    \begin{figure*}[t]
        \centering
        \includegraphics[width=0.75\linewidth]{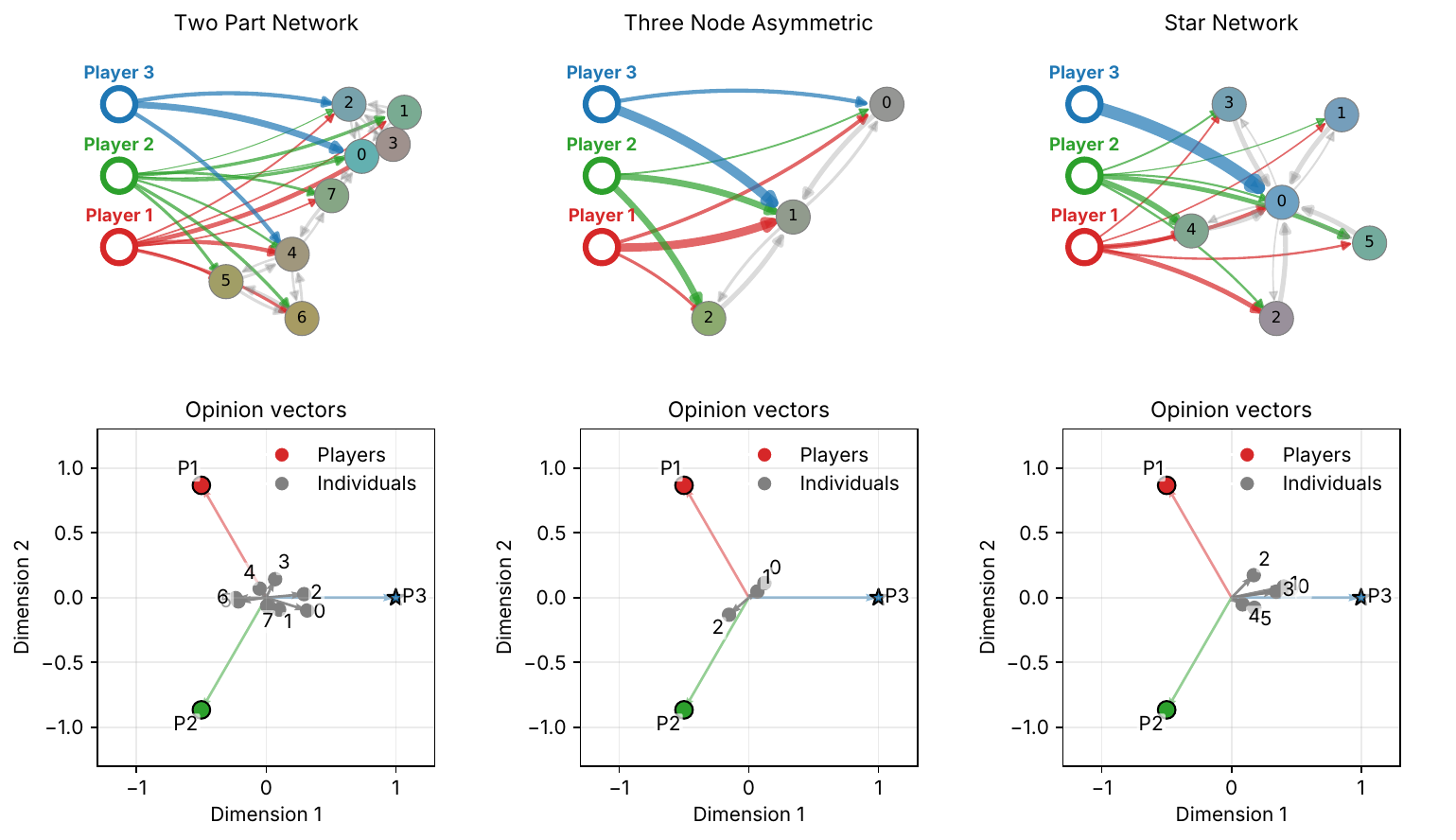}
         \vspace{-10pt}
        \small{\caption{The left side of the figure shows the network and players, along with the influence they allocate. Player three is optimized with the Iterated Linear solver and other players are random. The outlined and annotated circles are the players, and the solid color nodes are the individuals. The right side of the figure shows the reference opinions of each of the players in their colors, and the opinions of each individual in gray.}\label{fig:ex-all}}
         \vspace{-10pt}
    \end{figure*}

}{
    \begin{figure*}[t]
        \centering
        \includegraphics[width=0.75\linewidth]{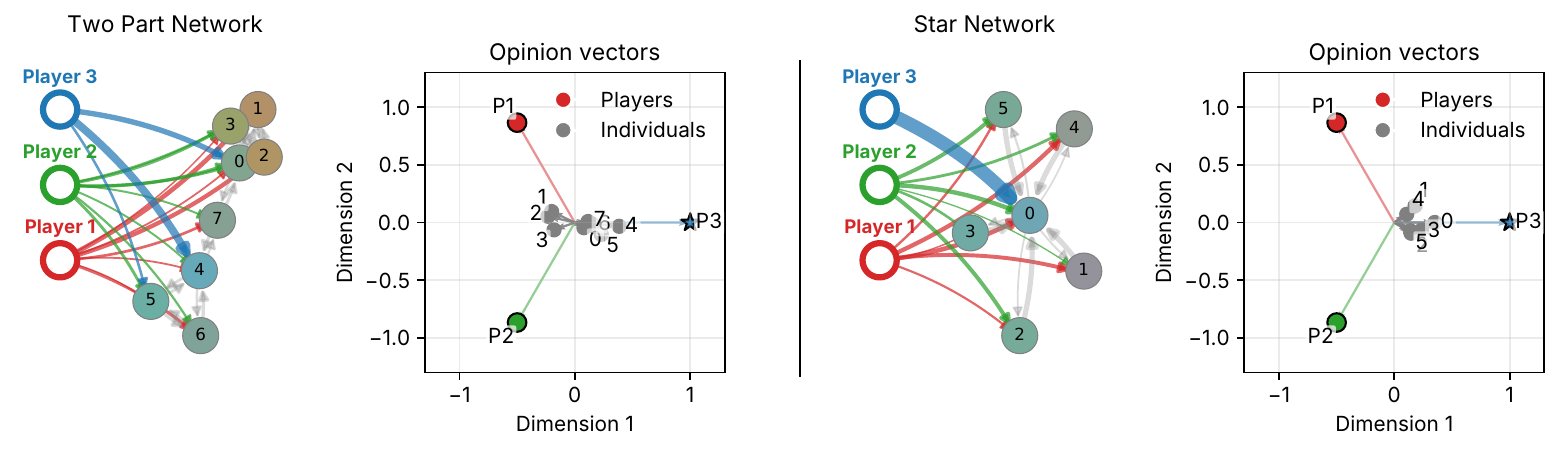}
         \vspace{-10pt}
        \small{\caption{Each figure shows a social network and the resulting influence allocation from \cref{sec:case-studies}. The left side of each figure shows the social network and the allocations of each player, and the right side of each figure shows the final opinions of each individual in the network. The optimized player is Player 3, shown in blue.
        }}
        \label{fig:ex-all}
         \vspace{-10pt}
    \end{figure*}
}

We aim to evaluate whether the proposed IL solver can achieve performance comparable to general solvers while maintaining scalability to large social networks. To assess this, we sample $3$ connected graphs for each $M \in [10, 50, 100, 200, 300, 400, 500, 600, 700, 800, 900, 1000]$ individuals according to the Erdős–Rényi model \cite{erdos_random_1958}, where the probability of an edge is $0.6$. The weights for each edge are then randomly sampled in order to yield a social network. The influence budget is fixed at $\lambda=0.5$. The initial opinions of the nodes are set to zero, and the reference opinions are set as described in Section \ref{sec:ref-obj}.The asymptotic state of the social network does not depend on the initial opinions and is set to a vector of zeros. The \cref{fig:example-topologies} shows examples of the resulting topologies. All opponents take random actions sampled from a uniform distribution.

We evaluated the performance of our proposed IL solver against three other solvers: a Non-Linear solver (NLP) \cite{andersson_casadi_2019}, a genetic solver (CMA-ES) \cite{hansen_cma-espycma_2023}, and a DCCP solver \cite{shen2016disciplinedconvexconcaveprogramming}. All solvers are given $2000$ seconds before they are terminated.

We measured the performance of each solver in terms of the objective value achieved and the computation time taken to reach that objective. For each setting, the influence budget is fixed at $\lambda = 0.5$. The genetic solver is GPU-accelerated, all other solvers are not. The runtimes are shown in \cref{fig:solver-runtime}. Missing data indicate that the solver did not finish in the allocated budget. Only our solver was able to produce solutions for $M=[900,1000]$. Our IL approach offers the lowest runtime for social networks over $50$ individuals of all tested approaches.

Objective improvement is presented in relation to a random baseline in \cref{tab:solver-perf}.
The NLP, DCCP, and IL solvers all achieve large increases in the objective when compared to the random baseline. As will be shown in \cref{sec:infl-budg-impact}, objective improvement trends downwards as the number of individuals increases. While the average performance of the NLP solver is highest, it is important to note that no solutions were produced for networks with over $100$ individuals. The genetic solver produces solutions in line with those of the IL and NLP solvers, but is unreliable, and occasionally produces solutions with large negative improvement. The DCCP solver improves over the baseline, but does not achieve the performance that the IL or NLP solvers achieve, likely due to the complexity of the constraint function in \cref{eq:dc-constraint}. The DC formulation also has larger decision variables, with the additional variable $z\in \mathbb{R}^{MD}$ requiring optimization. The IL solver achieves an average improvement over all social network sizes within $7\%$ of the NLP solver.

\iftoggle{arxiv}{
    \subsection{Archetypical Examples}\label{sec:case-studies}
    To complement random graphs, we analyze three prototype networks that highlight how equilibrium influence allocations depend on network topology. In a three-player scenario, we show the optimized actions of Player 3 against random actions taken by Players 1 and 2. The influence budget is fixed at $\lambda=1.0$. 
    
    \subsubsection{Three-Node Asymmetric Network}
    This social network is made up of three individuals, where the trust the central individual places in their neighbors is not symmetric. \cref{fig:ex-all} shows the resultant influence allocation and the final opinions of the individuals in the network. In line with our intuition, Player 3 allocates all of their influence to the central node.
    
    \subsubsection{Star Network}
    This social network has a central node to which all nodes are connected. 
    %The topology and results are shown in \cref{fig:ex-all}.  
    As shown in \cref{fig:ex-all}, the optimal strategy is to allocate all influence to the central node, which is the only node capable of disseminating influence to the entire network.
    %Due to the row-stochastic property required in the DeGroot model, each leaf node must allocate all of its influence to the center node. 
    % The optimized move for Player 3 allocates all the influence to the center node.
    A series of overlapping star networks could be used to represent social media influencers.
    
    \subsubsection{Two Connected Cliques}
    This social network consists of two fully connected cliques: one with five individuals and another with three individuals. These two cliques are connected by an intermediate or bridge individual who serves as the sole link between them.
    
    The optimized move for Player 3 involves allocating a majority of influence to the larger clique and a minority of influence toward the node in the smaller clique that is connected to the bridge. Notably, the optimal strategy does not include allocating influence to the bridge itself, but rather to the adjacent node in the other clique. This node is the highest degree node in the smaller clique.
}{
    \subsection{Archetypical Examples}\label{sec:case-studies}

    To complement random graphs, we analyze two prototype networks that highlight how equilibrium influence allocations depend on network topology. In a three-player scenario, we show the optimized actions of Player 3 against random actions taken by Players 1 and 2. The influence budget is fixed at $\lambda=1.0$.

    \subsubsection{Star Network}
    This social network has a central node to which all nodes are connected. 
    %The topology and results are shown in \cref{fig:ex-all}.  
    As shown in \cref{fig:ex-all}, the optimal strategy is to allocate all influence to the central node, which is the only node capable of disseminating influence to the entire network.
    %Due to the row-stochastic property required in the DeGroot model, each leaf node must allocate all of its influence to the center node. 
    % The optimized move for Player 3 allocates all the influence to the center node.
    A series of overlapping star networks could be used to represent social media influencers.
    
    \subsubsection{Two Connected Cliques}
    This social network consists of two fully connected cliques: one with five individuals and another with three individuals. These two cliques are connected by an intermediate or bridge individual who serves as the sole link between them.
    
    The optimized move for Player 3 involves allocating a majority of influence to the larger clique and a minority of influence toward the node in the smaller clique that is connected to the bridge. Notably, the optimal strategy does not include allocating influence to the bridge itself, but rather to the adjacent node in the other clique. This node is the highest degree node in the smaller clique.
}

\begin{figure}[h]
    \centering
    \vspace{-10pt}
    \includegraphics[width=0.9\linewidth]{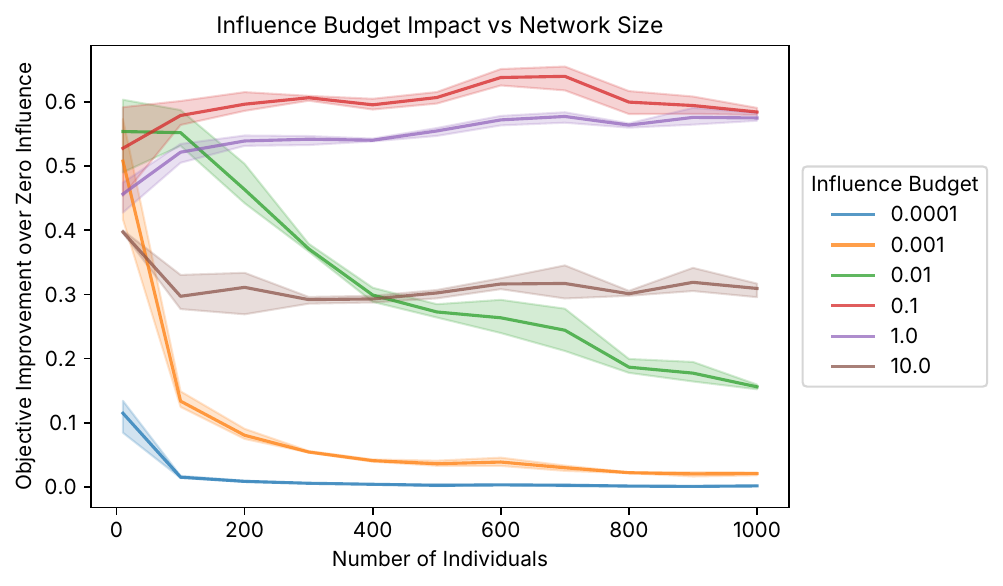} 
    \vspace{-10pt}
    \small{\caption{This figure shows how much the play of the Social Influence Game can distort the opinion of the network. The figure shows the relationship between the magnitude of the change and the number of individuals in the network, for different settings of $\lambda$. Larger influence budgets result in smaller distortions, as do larger social networks.}} 
    \label{fig:budget_vs_impact}
\end{figure}

\subsection{Influence Budget Impact}\label{sec:infl-budg-impact}
The influence budget allocated to players in the Social Influence Game has a substantial effect on the development of the asymptotic opinion. To evaluate this effect, we generated a series of random small-world networks using the Watts–Strogatz algorithm \cite{watts_collective_1998}, with the number of individuals $M \in [10,1000]$. For each network, we first established the baseline opinions in the absence of any influence agents, the Zero Influence scenario. The opinions of the other players were randomly selected, and the influence vector of the ego player was calculated with the Iterated Linear method of \cref{sec:iterated-linear}. The influence budget was varied from $\lambda=[0.0001,0.001,0.01,0.1,1.0,10.0]$, and the solutions were calculated in a three player setting.

The results, presented in \cref{fig:budget_vs_impact}, exhibit two trends. First, for small budgets, $\lambda \leq 0.01$, increases in the size of the social network decrease the objective improvement achieved. Second, increasing the influence budget has diminishing, then negative returns. While increasing the budget until $\lambda = 0.1$ increases the objective improvement, further increases yield negative returns. The objective improvement when $\lambda=1.0$ is lower than that where $\lambda=0.1$, and the objective improvement decreases sharply when $\lambda$ is further increased. At higher budgets, the relative impact of optimized allocation diminishes because the random actions of other agents represent a larger fraction of the total influence exerted within the network.

The strategic value of optimization is most pronounced when influence budgets are low, yet still adequate to effect change. In these settings, careful allocation of the influence budget can generate a significant change in the asymptotic opinion when faced with unsophisticated actors. In settings with high budgets, the actions of unsophisticated actors limit the impact of our optimized influence vector. Although the influence budget has a significant impact on the asymptotic opinion, as the size of the network increases, the play of the influence game generates a larger drift from the zero influence opinion.

    \section{Discussion}

Our results show that while the Social Influence Game can be posed as a DC program, directly applying DC solvers does not work well in practice. Our Iterated Linear solver leverages the structure of the problem and achieves solutions that are within $7\%$ of the Non-Linear solvers, up to $10$ times faster.

In archetypical examples, the solver consistently identifies the same structural leverage points that we would expect intuitively. Hubs receive the largest allocations, bridge nodes determine whether influence can flow between communities, and high-degree nodes amplify small amounts of budget. The alignment between the optimization results and network theory intuition suggests that the formulation effectively captures the underlying dynamics. Results from \cref{sec:iterated-linear} show that the relationship between centrality and budget allocation is not simply linear, but a piecewise linear relationship.

The budget scaling experiments show that the advantage generated by optimization is most pronounced in settings with low influence budgets. The trend between the overall change in opinion and the size of the network shows that a fixed influence budget has a greater impact on larger social networks.

    \section{Conclusion}

We introduced the Social Influence Game, a framework for modeling adversarial persuasion in social networks. We showed that the problem can be expressed as a DC problem, but that DCCP solvers are not as effective as our own Iterated Linear solver, which achieves performance close to a nonlinear solver with less required compute time.

We see this as a first step toward more realistic models of persuasion campaigns, where many actors compete simultaneously and resources must be deployed strategically. For future work, we aim to include consideration of changing network structure, finite-time objectives, system identification, nonlinear dynamics, and convergence guarantees.

    \bibliographystyle{IEEEtran-off}
\bibliography{references}
% \printbibliography
% \bibliography{IEEEabrv,references}

     \vspace{-5pt}
\appendix
\subsection{Opinion Bounds}
\begin{lemma} \label{lemma:opinion-bounds}
Given a social network which follows DeGroot dynamics, that is, that the opinions of the $i$-th individual at the next timestep $\x_i(t+1)$ are given by a convex combination of the other individuals in the network, the opinions will always remain in the convex hull of the initial opinions $\x_i(0)$ for all of the $M$ individuals in the network. Denoting the set of all initial opinions $X_0 = \{\x_i\quad \forall i \in [1, \ldots, M]\}$, we state:
 \vspace{-5pt}
\begin{equation}
    \x_i(t) \in \CVHull(X_0) \quad  \forall i\in [1,M], t
\end{equation}

\end{lemma}
\begin{IEEEproof}
    We prove this using induction. We know that the base case is true from the given, that all of the initial opinions must lie in their own convex hull at time $t=0$ by definition.
    \textbf{Inductive step:} We know that the system
    update is determined by a row-stochastic matrix $\W$. For a specific vertex,
    we can write the update as follows:
    \begin{align}
        \x(t+1)_{j} & = \sum_{k=0}^{M}\W_{jk}\x(t)_{k}
    \end{align}
    From the fact that $W$ is a row-stochastic matrix (from the definition of the DeGroot dynamics), we have
    \begin{gather}
        0 \leq W_{jk}\leq 1;\quad \sum_{k=0}^{n}W_{jk}= 1
    \end{gather}
    This shows that the opinion vectors at time $t+1$ are a convex combination of the previous opinion vectors, and are therefore within the convex hull of the opinion vectors at time $t$.
\end{IEEEproof}

\subsection{Proof of \cref{theorem:objective-equality}: Objective Equality}
\begin{IEEEproof}
First, we show that the minimizing opinion vectors for $J_1$ are the opinions which are equal to the reference opinion. We show that due to the invariance of the convex hull, this also maximizes the objective $J_2$.

\textbf{Optimization of $J_1$.}
The vector $\vec{v}$ that minimizes the expression $\|\R_i - \vec{v}\|_2$ is $\vec{v} = \R_p$ by the positive definite property of the norm, that the norm is zero if and only if the vector is zero. Recognizing that the expression $(\sum_{k=1}^M \x_k(t))/M$ is the average of the opinions of the $k$ individuals in the network, the minimizer at time $t$ is that where $\x_k(t) = \R_p \; \forall k \in [1, \ldots, M]$.

\textbf{Forward invariance of the convex hull.}
The players are modeled as individuals in the DeGroot network with trust placed completely in themselves. Due to \cref{lemma:opinion-bounds}, we know that they will always be in the convex hull of the original opinions. Because we stipulate that all opinions are within the convex hull of the reference opinions, we know that at all times, the opinions of players in the network are within the convex hull of the reference opinions.
\begin{equation}
     \x_i(t) \in \CVHull(\mathcal{R}) \quad  \forall i\in [1,M], t
\end{equation}
At all timesteps, we can define the opinion of an agent as a convex combination of the reference opinions in $\mathcal{R}$.

\textbf{Optimization of $J_2$.}
The expression $\left(\mathbb{1}_{M}^{\top}\otimes \I_{D}\right) \vec{x}_{m}(t)$ can be rewritten as $\sum_{k=1}^M \x_k(t)$. We can write this as $\sum_{k=1}^M \R_p^\top \x_k(t)$. Due to \cref{lemma:opinion-bounds}, we know we can rewrite each $\x_k(t) = \sum_{i=1}^{P}\gamma_i\R_i$, where $\sum_{i=1}^P\gamma_i = 1; \gamma_i \geq 0\; \forall i\in P$. We consider maximizing the norm of a single individual $\x_k(t)$ independently. We can write the dot product maximization as follows.
\begin{equation}
    \argmax_{\gamma}\R_p\x_k(t) = \sum_{i=1}^P \gamma_i\R_p^\top \R_i
\end{equation}
Cauchy-Schwarz gives the bound $\R_p^\top\R_i \leq \|\R_p\|\|\R_i\|$. Because the reference opinions when $\R_i \neq \R_p$ are not linearly dependent on $\R_p$, we get the strict inequality
$\R_p^\top \R_i < \| \R_p\|\| \R_i\| = 1$, and when $\R_i = \R_p$, $\R_p^\top \R_i = \|\R_p\|\|\R_i\| = 1$. Therefore, the component-wise maximum dot product we can get is $\R_p^\top \R_p$, so the maximizing choice of $\gamma_k=1; \gamma_i = 0\; \forall i\neq p$, which can be simplified to $\x_k(t) = \R_p$. Therefore, the opinion that maximizes the norm is $\x_k(t) = \R_p \; \forall k \in [1, \ldots, M]$.

\textbf{Equivalence.}
The minimizer of $J_1$ and the maximizer of $J_2$ are the same, the setting where $\x_k(t) = \R_p \; \forall k \in [1, \ldots, M]$, completing the proof.
\end{IEEEproof}
\end{document}